\documentclass[%
 reprint,
superscriptaddress,
 amsmath,amssymb,
 aps
]{revtex4-2}

\usepackage{soul}
\usepackage{xcolor}
\usepackage{braket}
\usepackage{graphicx}
\usepackage{dcolumn}
\usepackage{bm}

\usepackage{hyperref}
\hypersetup{
    colorlinks=true,
    linkcolor=blue,
    filecolor=magenta,      
    urlcolor=cyan,
    pdftitle={Logical error estimation from syndrome data of surface-code experiments},
    pdfpagemode=FullScreen,
    }
\begin{document}

\title{Logical error estimation from syndrome data of surface-code experiments}


\author{Evangelia Takou}
\email{evangelia.takou@duke.edu}
\affiliation{Duke Quantum Center, Duke University, Durham, NC 27701, USA}
\affiliation{Department of Electrical and Computer Engineering, Duke University, Durham, NC 27708, USA}

\author{Cesar Benito}
\affiliation{Instituto de Fisica Teorica UAM-CSIC, Universidad Autonoma de Madrid, Cantoblanco, 28049, Madrid, Spain}

\author{Arian Vezvaee}
\affiliation{Department of Electrical \& Computer Engineering, University of Southern California, Los Angeles, California 90089, USA}
\affiliation{Center for Quantum Information Science \& Technology, University of
Southern California, Los Angeles, CA 90089, USA}
\affiliation{Quantum Elements, Inc., Westlake Village, California, 91361, USA}


\author{Daniel A. Lidar}
\affiliation{Department of Electrical \& Computer Engineering, University of Southern California, Los Angeles, California 90089, USA}
\affiliation{Center for Quantum Information Science \& Technology, University of
Southern California, Los Angeles, CA 90089, USA}
\affiliation{Quantum Elements, Inc., Westlake Village, California, 91361, USA}
\affiliation{Department of Physics \& Astronomy, University of Southern California,
Los Angeles, California 90089, USA}
\affiliation{Department of Chemistry, University of Southern California,
Los Angeles, California 90089, USA}

\author{Kenneth R. Brown}
\affiliation{Duke Quantum Center, Duke University, Durham, NC 27701, USA}
\affiliation{Department of Electrical and Computer Engineering, Duke University, Durham, NC 27708, USA}
\affiliation{Department of Physics, Duke University, Durham, NC 27708, USA}
\affiliation{Department of Chemistry, Duke University, Durham, NC 27708, USA}

\date{\today}
\begin{abstract}
Decoders for quantum error correction (QEC) experiments rely on detector error models (DEMs), which encode, for each error, its probability and the detectors and logical observables it flips. Here we show that estimating DEM event probabilities from experimental syndromes is feasible, avoids independent device benchmarking, and produces useful decoder priors for estimating and reducing decoded logical error probabilities. We evaluate our methods using open-source data from surface-code memory experiments performed on Google's Willow chip, and we carry out analogous surface-code experiments on IBM's \texttt{ibm\_miami} processor. Despite the different physical error scales of the Google and IBM devices, in both cases our estimated DEMs improve logical error probabilities relative to baseline device-informed DEMs, typically at the $5\%-10\%$ level and with  larger gains in some IBM cases, without additional calibration circuits, decoder fine-tuning, or supervised fitting to logical outcomes.
\end{abstract}

\maketitle

In recent years, there has been rapid development in quantum error correction (QEC) with experiments demonstrating below threshold performance~\cite{GoogleNature2025, He2025PRL}, logical teleportation~\cite{MonzNature2021,AndersonScience2024}, and magic state distillation~\cite{RodriguezNature2025}.
Decoding algorithms have also been developed to improve the accuracy of QEC~\cite{BeniArXiv2025,BauschNature2024,SeniorArXiv2026,CaoArxiv2026} and reduce decoding times~\cite{mullerArXiv2025,YeArXiv2025}. A decoder can benefit from error models that accurately reflect the probabilities and detector-level signatures of hardware faults. 

Traditional characterization methods, such as tomography~\cite{ArianoPRL2001,Mohseni:2008ly}, can in principle provide detailed gate- or qubit-level error models, but they are resource-intensive and do not scale to the large space-time decoding problems relevant for QEC. Other Pauli noise-learning methods require pre-optimization of extra circuits that need to be run on the same device before the QEC experiment~\cite{HarperPRXQ2025}. These issues can be addressed by learning the detector-level model used by the decoder, namely the detector error model (DEM)~\cite{EisertarXiv2024} associated with the QEC circuit and its effective noise process (see Appendix~\ref{app:DEMIntro} for a brief review of DEMs that defines the terminology pertinent to our work). Several works have explored theoretically the idea of direct DEM estimation using syndromes in QEC ~\cite{YoungArxiv2025,BrownPRA2025,BrownArXiv2025,BhardwajArXiv2025,HinesArXiv2026}. Recently, Ref.~\cite{UlrichArXiv2026} consolidated several DEM-estimation algorithms and applied them to Google’s memory experiment data~\cite{GoogleNature2025}, after first validating rate and structure recovery on simulated syndromes, an aspect that had previously been explored for various codes and noise models in Refs.~\cite{BrownPRA2025,BrownArXiv2025}. However, decoder-level performance from experimental syndromes was explored only to a limited extent, with raw logical errors reported for a single case: the $d=7$, 10-cycle $X$-memory surface code. Based on this case alone, it remains unclear whether syndrome-estimated DEMs systematically improve logical performance relative to DEMs constructed from average device performance or more refined approaches such as reinforcement learning (RL)~\cite{NewmanPRL2024}.

This leaves the following question unresolved: can DEM event probabilities learned directly from the syndrome history of a QEC experiment, on a fixed reference detector-logical support, serve as a competitive decoder prior without separate calibration circuits, detailed device benchmarking of those probabilities, or supervised decoder-prior fitting? Here, we answer this question by analyzing the logical performance of DEMs estimated from experimental syndromes of surface-code memory experiments on Google's Willow chip and IBM's \texttt{ibm\_miami} processor. We show that these syndrome-estimated DEMs are reliable decoder priors and reduce logical error probabilities relative to baseline DEMs.
For the Willow surface-code data, our estimated DEMs perform comparably to DEMs obtained from reinforcement learning optimization. For the IBM data, 
the estimated DEMs reduce the decoded logical error probability relative to decoding with an IBM-like baseline DEM, with the largest fractional reduction
of $\sim 38\%$ occurring for a single syndrome-extraction cycle. These demonstrations show that syndrome data, combined with a reference detector-logical support, can provide useful decoder priors across different devices, code distances, logical states, and numbers of syndrome-extraction cycles.


\begin{figure*}[!htbp]
    \centering
    \includegraphics[scale=0.746]{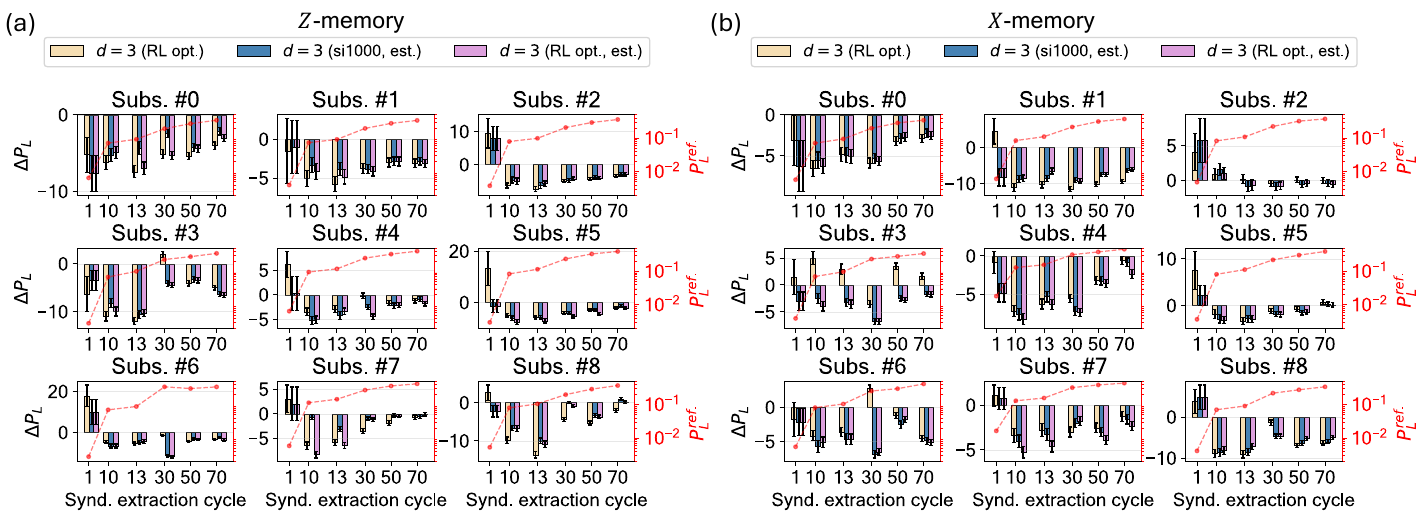}
    \caption{Logical error probability comparison for the $d=3$ surface-code memory experiments on Google Willow. (a) Fractional percent change in LEP for $Z$-memory experiments versus syndrome (synd.) extraction cycle. Each panel corresponds to a different subsystem (subs.) of the device. (b) Same as in (a) for $X$-memory experiments. All bars are plotted using 
$\Delta P_L(A|B)$ [Eq.~\eqref{eq:DeltaP_L}], with the SI1000 DEM used as the reference. The yellow bars show $\Delta P_L(\mathrm{RL}|\mathrm{SI1000})$, where $\mathrm{RL}$ denotes the reinforcement-learning optimized model (RL opt.). The blue bars show $\Delta P_L(\mathrm{est}_{\mathrm{SI1000}}|\mathrm{SI1000})$, where $\mathrm{est}_{\mathrm{SI1000}}$ denotes the DEM whose rates are estimated from syndrome data using the SI1000 detector-logical support and the same hyperedge decomposition as the SI1000 DEM. The pink bars show $\Delta P_L(\mathrm{est}_{\mathrm{RL}}|\mathrm{SI1000})$, where $\mathrm{est}_{\mathrm{RL}}$ denotes the DEM whose rates are estimated from syndrome data using the RL-optimized detector-logical support and its corresponding hyperedge decomposition. 
Negative values indicate lower LEP than the SI1000 reference.  Error bars show the delta-method standard error of $\Delta P_L$, obtained by propagating the binomial variances of the two LEP estimates being compared and their covariance, since both LEPs are evaluated on the same shot ensemble. The red dashed curve with markers, plotted against the red right-hand axis, shows the SI1000 reference logical error probability $P_L^{\mathrm{SI1000}}$ used in the denominator of all plotted $\Delta P_L$ values.}
    \label{fig:Google_d_3}
\end{figure*}


\textit{Estimations from Willow processor.} To test our estimation method, we use  syndrome data~\cite{DataFromGoogle} from the rotated surface code experiments of Ref.~\cite{GoogleNature2025}. Together with the syndrome history of $50,000$ shots for $d\in\{3,5,7\}$ $X$- and $Z$-memory experiments, these data include two reference DEMs. The first is the SI1000 DEM, generated via Stim~\cite{Gidney2021Stim} from the circuit-level superconducting noise model supplied with the Google data set. This model provides an experiment-informed but not syndrome-fitted decoder prior. The second is the reinforcement learning (RL) optimized DEM of Ref.~\cite{NewmanPRL2024}, whose decoder priors were optimized to improve logical performance. These two reference DEMs differ both in their error probabilities and in the way hyperedges are decomposed into graphlike components. 
In the data sets considered here, hyperedges appear for the multi-cycle cases, $r>1$.
Our goal is to compare the performance of these DEMs with that of DEMs estimated from syndromes. 
For each Willow experimental instance, we estimate the DEM event probabilities by averaging the relevant detector moments over the same $50,000$ publicly released experimental shots that are subsequently decoded. The estimation step uses only detector-event statistics and the detector-logical support inherited from the chosen reference DEM; it does not use the final logical success or failure outcomes. Since we do not average over time-translated detector locations, these estimated DEM event probabilities can vary from one syndrome-extraction cycle to the next.  An alternative approach would be to estimate a time-averaged DEM (see Appendix~\ref{app:Rates}). As we show below, even without time-averaging, the resulting DEMs provide useful decoder priors and can reduce the decoded logical error probability. We assess logical performance by decoding the experimental syndrome data with Minimum-Weight Perfect Matching (MWPM)~\cite{HiggotACM2022,HiggotQuantum2025}, using each DEM as the decoder prior. This yields a logical error probability (LEP), denoted by $P_L$, for each DEM. Although LEP is sometimes described as a \textit{logical error rate} in the QEC literature, we use the term logical error probability throughout, since the quantity reported here is a failure probability for a finite memory experiment with a specified number of syndrome-extraction cycles.
To compare the performance of two DEMs, we use the fractional percent change, 
\begin{equation}
\label{eq:DeltaP_L}
\Delta P_L(A|B)=\frac{P_L^{A}-P_L^{B}}{P_L^{B}}\times 100\%,
\end{equation}
where $B$ is the reference DEM. 
Negative $\Delta P_L(A|B)$ 
means that DEM $A$ yields a smaller LEP than the reference DEM $B$.
For the Willow experiments, 
the reference is the SI1000 DEM, so $B=\mathrm{SI1000}$ and $A$ is either the RL-optimized DEM
or our estimated DEM.

\begin{figure*}[!htbp]
    \centering
    \includegraphics[scale=0.8]{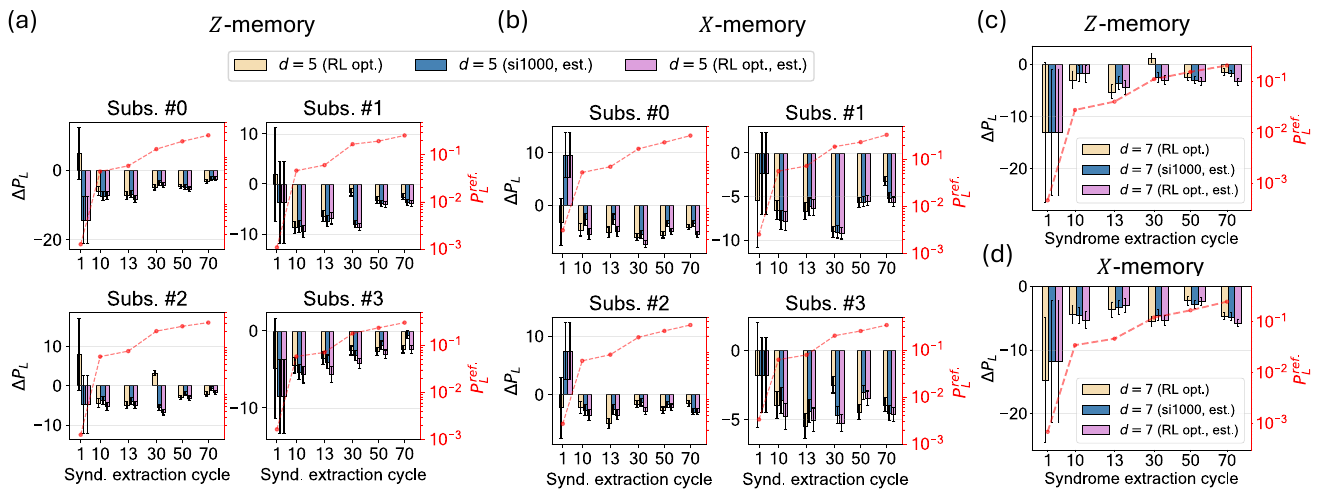}
    \caption{Logical error probability comparison for the $d=5$ and $d=7$ surface-code memory experiments on Google Willow. Panels (a),(b) show the fractional percent change in LEP for the $d=5$ $Z$- and $X$-memory experiments, respectively, versus syndrome-extraction cycle; each subpanel corresponds to a different subsystem. Panels (c),(d) show the corresponding $d=7$ $Z$- and $X$-memory results. As in Fig.~\ref{fig:Google_d_3}, all bars are plotted using $\Delta P_L(A|B)$ [Eq.~\eqref{eq:DeltaP_L}] with $B=\mathrm{SI1000}$. The yellow, blue, and pink bars show $\Delta P_L(\mathrm{RL}|\mathrm{SI1000})$, $\Delta P_L(\mathrm{est}_{\mathrm{SI1000}}|\mathrm{SI1000})$, and $\Delta P_L(\mathrm{est}_{\mathrm{RL}}|\mathrm{SI1000})$, respectively, with the same definitions as in Fig.~\ref{fig:Google_d_3}. Error bars and the red dashed curve are also as in Fig.~\ref{fig:Google_d_3}.}
    \label{fig:Google_d_5_Z}
\end{figure*}



The $d=3$ surface-code memory experiments in Ref.~\cite{GoogleNature2025} were run on nine different subsystems on the device, where each subsystem is a distinct distance-three surface-code patch embedded in the Willow processor. Our results for $\Delta P_L$ are summarized in Fig.~\ref{fig:Google_d_3},  for each subsystem, memory experiment ($Z$ or $X$), and number of syndrome-extraction cycles $r\in\{1,10,13,30,50,70\}$. The yellow bars report $\Delta P_L(\mathrm{RL}|\mathrm{SI1000})$. The blue bars correspond to $A=\mathrm{est}_{\mathrm{SI1000}}$: a DEM whose rates are estimated from syndrome data using the SI1000 detector-logical support and the same hyperedge decomposition used for the SI1000 DEM. 
The pink bars correspond to $A=\mathrm{est}_{\mathrm{RL}}$: a DEM whose rates are estimated from syndrome data using the RL-optimized detector-logical support, which has a different hyperedge decomposition. (For $r=1$, the two estimated DEMs are identical in our construction because the corresponding reference DEMs have the same graphlike detector-logical support; the only distinction between the two constructions, namely how hyperedges are decomposed, is absent in this case.)

Overall, the syndrome-estimated DEMs usually improve on the SI1000 reference, i.e., they give negative $\Delta P_L(A|\mathrm{SI1000})$, with LEP reductions of up to $\sim 10\%$ in some cases. Representative examples include the $Z$-memory data for subsystem $\#6$ at $r=30$, subsystem $\#8$ at $r=13$, and subsystem $\#3$ at $r=13$. The syndrome-estimated DEMs also generally perform comparably to the RL-optimized DEM: in some cases $\Delta P_L(\mathrm{est}_{\mathrm{SI1000}}|\mathrm{SI1000})$ or $\Delta P_L(\mathrm{est}_{\mathrm{RL}}|\mathrm{SI1000})$ is more negative than $\Delta P_L(\mathrm{RL}|\mathrm{SI1000})$, while in other cases the RL-optimized DEM gives the larger reduction. A noteworthy feature is that the RL-optimized DEM sometimes underperforms the SI1000 DEM, giving positive $\Delta P_L(\mathrm{RL}|\mathrm{SI1000})$; examples include subsystem $\#5$ for both $Z$- and $X$-memory, subsystem $\#3$ for $X$-memory, and subsystem $\#6$ for both $Z$- and $X$-memory. In several of these cases, the syndrome-estimated DEMs are more robust: they either still outperform SI1000, as for subsystem $\#3$ in the $X$-memory data, or they increase the LEP by less than the RL-optimized DEM, as for subsystem $\#5$ in the $X$-memory data. The main exceptions are subsystems $\#2$ and $\#8$ for $X$-memory during some of the initial cycles.

We also note that the fractional gains decrease when $r\gg d$, as logical failures accumulate and the LEP approaches its saturation regime; 
in this regime the decoder prior has a smaller effect on the final logical decision. For $r=1$, the error bars on $\Delta P_L$ are larger because $P_L^{\mathrm{SI1000}}$ is small and appears in the denominator of the fractional change.

In Fig.~\ref{fig:Google_d_5_Z},  we show $\Delta P_L$ for the $d=5$ and $d=7$ memories.  The trends are consistent with the $d=3$ results of Fig.~\ref{fig:Google_d_3}: the syndrome-estimated DEMs remain competitive with the SI1000 and RL-optimized references as the code distance is increased, and in some cases yield LEP reductions of up to $\sim 10\%$ relative to SI1000.

We also compare our scope with that of Ref.~\cite{UlrichArXiv2026}. That work reported decoder performance for a limited logical-memory benchmark, including the raw number of logical failures for the $d=7$, $r=10$, $X$-memory case, and found comparable MWPM performance between the RL-optimized DEM and their syndrome-estimated DEM within error bars. Here we perform a systematic benchmark across all available subsystems, both memory basis states, and all available code distances and cycle counts. In several cases the syndrome-estimated DEMs outperform the RL-optimized DEMs, suggesting that direct syndrome estimation can provide competitive decoder priors without the training cost of RL optimization. The comparison using $\mathrm{est}_{\mathrm{SI1000}}$ tests syndrome rate estimation without RL-derived support, whereas the comparison using $\mathrm{est}_{\mathrm{RL}}$ tests how much performance is obtained by replacing the RL-optimized probabilities with syndrome-estimated probabilities while keeping the RL-derived detector-logical support and hyperedge decomposition fixed. Thus, the results distinguish rate estimation from the choice of DEM support and hyperedge decomposition. The estimation algorithms also differ: Ref.~\cite{UlrichArXiv2026} uses a distinct moment- and polarization-based framework, whereas our method estimates rates on the DEM supports and hyperedge decompositions described above.

\begin{figure}[]
    \centering
    \includegraphics[scale=1.2]{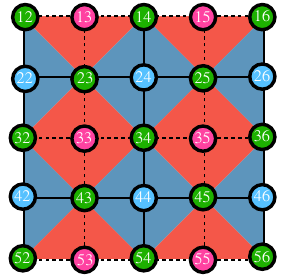}
    \caption{The unrotated $d=3$ surface code on \texttt{ibm\_miami} is arranged on a $5\times 5$ grid, with the upper-left qubit identified as qubit $12$ on the \texttt{ibm\_miami} device. Data qubits (green circles) lie along the edges of the square lattice (black lines). Plaquette regions (blue) define $Z$-type stabilizers, while vertex regions (red) define $X$-type stabilizers. These stabilizers are measured simultaneously by transferring syndrome information to bare ancilla qubits (red and blue circles) via the dashed and solid syndrome–data couplings. In contrast, Ref.~\cite{GoogleNature2025} implemented the rotated form of the ZXXZ surface code~\cite{Bonilla-Ataides:2021aa} on the Willow chip.}
    \label{fig:surface-code-ibm}
\end{figure}

\textit{Estimations from \texttt{ibm\_miami}.} We now explore the performance of our method on \texttt{ibm\_miami}. We run $d=3$ unrotated surface-code memory circuits, shown in Fig.~\ref{fig:surface-code-ibm}, with up to $r=19$ syndrome-extraction cycles and perform $X$- and $Z$-memory experiments. Because unconditional qubit reset is not implemented in the device~\cite{harper2025arxiv}, we do not reset ancilla qubits between cycles. This requires a modified detector definition, in which a measurement outcome is compared with the corresponding outcome from two cycles earlier~\cite{vezvaee2025arxiv}. Moreover, the scheduled two-qubit gates 
on \texttt{ibm\_miami} have varying durations, which creates idle gaps in the circuits. We fill these gaps with XY4 dynamical decoupling (DD) sequences~\cite{Gullion:1990aa,Viola:99}, providing an additional layer of error suppression that has been shown to be compatible with QEC~\cite{Vezvaee2026natcomm,kasatkin2026arxiv}. 

These aspects of our \texttt{ibm\_miami} experiments contrast with those of Ref.~\cite{GoogleNature2025}: we use a no-reset approach~\cite{geher2025arxiv}, and we apply DD to all circuit gaps, rather than only during reset and readout. 


\begin{figure}[!htbp]
    \centering
    \includegraphics[scale=0.89]{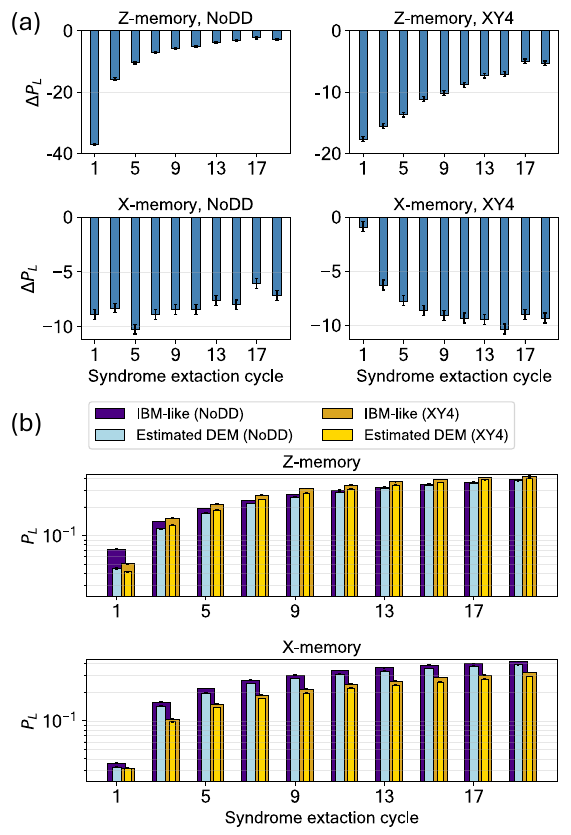}
    \caption{Logical performance on \texttt{ibm\_miami} for the $d=3$ unrotated surface code. (a) Fractional percent change in LEP, $\Delta P_L=(P_L^{\mathrm{est_{IBM}}}-P_L^{\mathrm{IBM\mbox{-}like}})/P_L^{\mathrm{IBM\mbox{-}like}}\times 100\%$, versus syndrome extraction cycle. Each panel shows either $Z$- or $X$-memory, without DD or with XY4. Negative values indicate an LEP reduction for the estimated DEM relative to the IBM-like DEM. (b) Corresponding logical error probabilities, $P_L$, for the IBM-like and estimated DEMs. Error bars show the standard deviation.}
    \label{fig:IBM_miami}
\end{figure}

We follow a similar procedure as for Willow, comparing a Stim-generated IBM-like DEM with a syndrome-estimated DEM. The IBM-like DEM is generated using the Pauli circuit-level noise model described in Appendix~\ref{app:ibm-like-noise}. To construct the syndrome-estimated IBM DEM ($A=\mathrm{est_{IBM}})$, we take the undecomposed detector-logical supports of the IBM-like DEM and estimate the corresponding DEM event probabilities from the experimental syndrome data. We then apply the same graphlike hyperedge decomposition before decoding with MWPM. Thus, both the IBM-like and syndrome-estimated DEMs are decoded with MWPM after graphlike decomposition. In Fig.~\ref{fig:IBM_miami}(a) we show $\Delta P_L(\mathrm{est_{IBM}}|\mathrm{IBM\mbox{-}like})$, for $Z$- and $X$-memory surface-code experiments, both without DD and with XY4. 
Across all cases, the LEP reduction is of order $5\%$-$10\%$, with the largest reduction recorded for $r=1$ in the $Z$-memory data ($\sim 37\%$ without DD, and $\sim 18\%$ with XY4). The IBM-like DEM is not calibrated to the effective idle-noise suppression produced by DD; it uses the same Pauli circuit-level assumptions with or without DD, aside from the gates and idle intervals present in the compiled circuit. Thus the IBM-like LEP for the $Z$-memory can be worse with DD than without DD.  The syndrome-estimated DEM, by contrast, is inferred from the syndrome data and therefore captures effective DEM event probabilities with or without DD, reducing the LEP in both cases compared to the IBM-like DEM.
In Appendix~\ref{ibm_miami_Corr_MWPM}, we also decode the DEMs with correlated MWPM, and find larger LEP reductions.

During the DEM event-probability estimation step, the formulas of Appendix~\ref{app:Rates} use empirical detector moments and covariances. For the $Z$-memory without DD, we found that some of these quantities are negative and statistically significant; the corresponding diagnostic data are shown in Appendix~\ref{app:ibm-miami_extra}. These negative correlators are associated with a detector family whose members fire with probability greater than $0.5$ at time-translated locations. Such behavior is consistent with a large coherent or biased error affecting the associated ancilla or its neighboring couplings, although the detector firing probability alone is not sufficient to identify the microscopic error mechanism. Coherent ZZ crosstalk is a significant noise source on \texttt{ibm\_miami}~(see Appendix~\ref{app:Crosstalk}), and is one plausible mechanism because it is suppressed by DD~\cite{Tripathi2022PRApp,Zeyuan:22,Vezvaee2026natcomm}.
A large marginal detector firing probability can produce negative covariances with other detectors when the corresponding coincidence rates are smaller than the product of the marginals. Although negative correlators do not automatically invalidate every DEM event probability involving this detector family, they can lead to unphysical DEM event-probability estimates. In those cases we set the corresponding DEM event probability to zero, as described in Appendix~\ref{app:Rates}.

This limitation is consistent with the behavior of $\Delta P_L$: for the $Z$-memory without DD shown in Fig.~\ref{fig:IBM_miami}(a, top left), the gains saturate and are not appreciable for the largest $r$ values considered. When XY4 DD is included in Fig.~\ref{fig:IBM_miami}(a, top right), the large coherent error is suppressed and we do not encounter negative correlators; the gains are correspondingly more pronounced for the largest numbers of cycles. For the $X$-memory, the same time-translated detector fires with probability $\sim 45\%$, giving rise to negative covariances of order $10^{-4}-10^{-3}$ in fewer locations than in the $Z$-memory without DD. For the $X$-memory with XY4, these negative covariances are eliminated. Thus, negative correlations are not a limiting issue for the $X$-memory (see Appendix~\ref{app:ibm-miami_extra}), and the $X$-memory with XY4 has the lowest LEP among the IBM memory experiments considered here, as seen in Fig.~\ref{fig:IBM_miami}(b). 

A second limitation is that the hierarchical subtraction procedure of Appendix~\ref{app:Rates} yields negative estimates for some single-detector boundary DEM event probabilities, i.e., probabilities assigned to DEM events that toggle only one detector. We set the corresponding DEM event probabilities to zero. 
This issue occurs in the IBM data but not in the Google data, possibly because the Google circuits include leakage removal mechanisms and use ancilla reset, whereas the IBM experiment does not reset ancillas. Despite this limitation, decoding with the partially estimated DEM still reduces the observed logical error probability relative to the baseline DEMs considered here.

Additional comparisons using the SPAM-normalized entanglement-fidelity metric of Ref.~\cite{vezvaee2025arxiv} are provided in Appendix~\ref{app-EF}.

\textit{Conclusions and discussion.} We presented a noise-learning method that uses only the syndrome history of QEC experiments. DEM-based noise estimation retains detector-level information directly relevant for decoding, without requiring separate characterization circuits or a detailed microscopic device model. We benchmarked the method on Google and IBM data and obtained lower logical error probabilities than baseline device-informed DEMs. The method exploits spatial and temporal inhomogeneities and can capture detector-level signatures of non-Pauli physical noise under the experimental operating conditions. A notable feature in both devices is that LEP improvements are obtained even when rates are estimated only on the DEM support generated by stochastic circuit-level noise. Additional DEM events may exist because of coherent or correlated errors. We did not test whether adding such events would further improve performance, in order to benchmark the simplest version of the method rather than report fine-tuned results. 

A natural future direction is to test our methods beyond memory experiments, e.g., when logical gates are applied and followed by several syndrome-extraction cycles. It would also be useful to determine whether adding extra hyperedge events to the DEM, or using alternative hyperedge decompositions, can improve logical error suppression. Finally, identifying the DEM signatures and locations of leakage or loss events, and learning these signatures directly from syndromes, would be especially valuable for superconducting and neutral-atom QEC devices.


\acknowledgements{
The authors are grateful to Alejandro Bermudez for helpful discussions. All authors acknowledge support by the Office of the Director of National Intelligence (ODNI), Intelligence Advanced Research Projects Activity (IARPA), under the Entangled Logical Qubits program through Cooperative Agreement Number W911NF-23-2-0216. E.T. and K.R.B. acknowledge support by the ARO/LPS QCISS program (W911NF-21-1-0005). C.B. acknowledges support from the Spanish Ministry of Science, Innovation and Universities under grant FPU24/01105, from PID2024-161474NB-I00 (MCIU/AEI/FEDER,UE), from QUITEMAD-CM TEC-2024/COM-84, from the Grant IFT Centro de Excelencia Severo Ochoa CEX2020-001007-S, funded by MCIN/AEI/10.13039/501100011033, and from the CSIC Research Platform on Quantum Technologies PTI-001. For A.V. and D.L., this material is based upon work supported by, or in part by, the U. S. Army Research Laboratory and the U.S. Army Research Office under contract/grant number W911NF2310255. The experiments on IBM machines were conducted using IBM Quantum Systems provided through the University of Southern California's IBM Quantum Innovation Center. The views expressed are those of the authors and do not reflect the official policy or position of IBM or the IBM Quantum team.
}

\clearpage
\appendix

\section{Detector error models}
\label{app:DEMIntro}

This appendix gives a brief, self-contained introduction to detector error models (DEMs), which provide the classical probabilistic interface between a noisy QEC experiment and a decoder. A DEM is a stochastic model for the detector events and logical-observable flips that are relevant for decoding. DEMs are especially natural for repeated stabilizer-measurement experiments, such as surface-code memories, where the decoder does not observe the underlying physical faults directly. Instead, it receives a binary pattern of detector events derived from the syndrome-measurement history~\cite{Dennis:02,Fowler:2012ys}. The DEM formalism used in modern surface-code simulations is closely associated with Stim \cite{Gidney2021Stim}, and DEMs are now a standard representation for supplying decoder priors in experimental and simulated QEC studies \cite{GoogleNature2025,NewmanPRL2024,BauschNature2024}.

We begin with the distinction between raw measurement outcomes and detectors. A QEC experiment produces a record of raw measurement outcomes $m_a\in \mathbb{F}_2=\{0,1\}$. A detector is a specified parity of such outcomes whose value is deterministic in the noiseless reference circuit. Detector $i$ is defined by
\begin{equation}
\label{eq:DetectorDefinition}
    v_i^{\mathrm{exp}}=b_i\oplus\bigoplus_{a\in D_i}m_a ,
\end{equation}
where $D_i$ is the set of measurement outcomes entering the detector, $b_i$ is the noiseless reference value of the parity, and $\oplus$ denotes addition modulo two. The detector itself is the parity check specified by Eq.~(\ref{eq:DetectorDefinition}). A \emph{detector event} occurs when $v_i^{\mathrm{exp}}=1$, signaling a deviation from the noiseless reference behavior.

For example, in a repeated surface-code memory with reset, a detector is often the parity of two consecutive measurements of the same stabilizer,
\begin{equation}
    v_{j,t}^{\mathrm{exp}}=m_{j,t-1}\oplus m_{j,t},
\end{equation}
possibly with a known noiseless offset absorbed into $b_i$. If the stabilizer value is unchanged and both measurements are correct, then $v_{j,t}^{\mathrm{exp}}=0$. A data-qubit error occurring between rounds can change nearby stabilizer values and create a pair of nearby detector events in space-time. A measurement error can instead create a pair of detector events separated in time. Near an initial, final, or spatial boundary, an error can lead to a single detector event. In the no-reset IBM experiments we consider in this work, the detector definition is modified so that an ancilla measurement outcome is compared with the corresponding outcome from two cycles earlier, but the basic object is still a deterministic parity check of raw measurement outcomes in the noiseless circuit.

In the independent-event representation we use in this work, a DEM is specified by a finite set $\mathcal{M}=\{e\}$ of DEM events $e$, also called elementary error mechanisms. The label $e\in\mathcal{M}$ indexes one elementary Bernoulli error mechanism in the DEM. Each event $e$ is associated with the data
\begin{equation}
\label{eq:e-map}
    e\mapsto (p_e,\mathbf{h}_e,\boldsymbol{\ell}_e).
\end{equation}
Here $p_e$ is the probability assigned to event $e$, $\mathbf{h}_e$ is the detector-flip vector, and $\boldsymbol{\ell}_e$ is the logical-observable flip vector. We refer to the pair $(\mathbf{h}_e,\boldsymbol{\ell}_e)$ as the \emph{detector-logical signature} of event $e$.

In more detail, $p_e$ is the probability with which DEM event $e$ is sampled in the independent-event DEM. The way $p_e$ is obtained depends on how the DEM is constructed. In a circuit-generated DEM, a circuit-level noise model assigns probabilities $r_f$ to elementary physical fault locations $f$; each fault is propagated through the noiseless stabilizer circuit to determine the detectors and logical observables it flips. 
Independent faults with the same detector-logical signature as event $e$ are then combined by the parity rule
\begin{equation}
\label{eq:parity-rule}
    p_e=
    \frac{1-\prod_{f\in G_e}(1-2r_f)}{2},
\end{equation}
where $G_e$ is the set of physical faults whose detector-logical signature is the same as that of $e$. In the syndrome-estimated DEMs used in this work, the allowed detector-logical signatures are taken from a reference DEM, while the corresponding probabilities are inferred from empirical detector moments, as described in Appendix~\ref{app:Rates}. In a decoder-prior optimized DEM, such as the RL-optimized DEM, the probabilities should be understood as optimized decoder priors rather than as arising from hardware-calibration estimates.

The vector $\mathbf{h}_e\in\mathbb{F}_2^M$ is the detector-flip vector, where $M$ is the number of detectors and
\begin{equation}
\label{eq:h_ei}
    h_{e,i}=
    \begin{cases}
    1, & \text{if event } e \text{ toggles detector } i,\\
    0, & \text{otherwise.}
    \end{cases}
\end{equation}
Equivalently, $\mathbf{h}_e$ is the binary indicator vector of the detector support set
\begin{equation}
\label{eq:EeDefinition}
    E_e=\{i:h_{e,i}=1\}.
\end{equation}
Thus $E_e$ is the set of detectors toggled by event $e$, while $\mathbf{h}_e$ is the corresponding binary indicator vector. Note that two sampled DEM events that both toggle the same detector cancel in that detector bit. 

Similarly, $\boldsymbol{\ell}_e\in\mathbb{F}_2^K$ is the logical-observable flip vector, where $K$ is the number of tracked logical observables and
\begin{equation}
    \ell_{e,k}=
    \begin{cases}
    1, & \text{if event } e \text{ flips logical observable } k,\\
    0, & \text{otherwise.}
    \end{cases}
\end{equation}
Equivalently, $\boldsymbol{\ell}_e$ is the binary indicator vector of the logical-observable support set
\begin{equation}
    L_e=\{k:\ell_{e,k}=1\}.
\end{equation}
The vector $\boldsymbol{\ell}_e$ records the logical class of the error, for example whether event $e$ flips a tracked logical $X$ or logical $Z$ observable of a memory experiment. Equivalently, the detector-logical signature can be written as the pair of support sets $(E_e,L_e)$.

A DEM event is an elementary error only at the detector-model level. The DEM does not further resolve it into microscopic physical errors in the circuit or hardware. Such an event may be the detector-level image of a single circuit-level error, such as a Pauli error after a gate or a measurement error. 
For example, a DEM event that toggles detectors $i$ and $j$ and also flips logical observable $k$ has
$\mathbf{h}_e=\hat{\mathbf{e}}_i^{(M)}\oplus\hat{\mathbf{e}}_j^{(M)}$ and $\boldsymbol{\ell}_e=\hat{\mathbf{e}}_k^{(K)}$,
or equivalently $E_e=\{i,j\}$ and $L_e=\{k\}$. Here $\hat{\mathbf{e}}_i^{(M)}$ and $\hat{\mathbf{e}}_k^{(K)}$ denote standard basis vectors in the detector and logical-observable spaces, respectively. 

Sampling the DEM means independently drawing a Bernoulli random variable $x_e$ for each DEM event $e\in\mathcal{M}$, with
\begin{equation}
    \Pr(x_e=1)=p_e,
    \quad
    \Pr(x_e=0)=1-p_e.
\end{equation}
The sampled detector-event record and logical-flip vector are then
\begin{equation}
\label{eq:DEMSampling}
    \mathbf{v}^{\mathrm{DEM}}=\bigoplus_{e\in\mathcal{M}} x_e\mathbf{h}_e,
    \quad
    \boldsymbol{\lambda}^{\mathrm{DEM}}=\bigoplus_{e\in\mathcal{M}} x_e\boldsymbol{\ell}_e .
\end{equation}
Equation~(\ref{eq:DetectorDefinition}) and the component $v_i^{\mathrm{DEM}}=\bigoplus_{e\in\mathcal{M}} x_e h_{e,i}$ of Eq.~(\ref{eq:DEMSampling}) define the same kind of binary variable, namely the event variable associated with detector $i$, but at two different levels of description. In the experiment, $v_i^{\mathrm{exp}}$ is computed from the raw measurement record by taking the parity specified in Eq.~\eqref{eq:DetectorDefinition}. In a DEM sample, $v_i^{\mathrm{DEM}}$ is generated without simulating the raw measurement outcomes $m_a$. Instead, each sampled DEM event toggles detector $i$ if and only if $h_{e,i}=1$, as specified in Eq.~\eqref{eq:h_ei}, and the resulting detector bit is the parity of all such toggles.

In experimental decoding, the decoder is given the detector-event vector $\mathbf{v}^{\mathrm{exp}}$ computed from the measured syndrome record. The DEM supplies the decoder with a prior distribution over DEM events that could have produced such a record. By contrast, $\mathbf{v}^{\mathrm{DEM}}$ is the random detector-event vector obtained by sampling the DEM itself. A good DEM is one for which the distribution of $\mathbf{v}^{\mathrm{DEM}}$ approximates the empirical distribution of $\mathbf{v}^{\mathrm{exp}}$, together with the corresponding logical-flip information.

The decoder outputs an \emph{inferred} logical flip $\hat{\boldsymbol{\lambda}}(\mathbf{v}^{\mathrm{exp}})$, or a stabilizer-equivalent correction. In an experimental shot, the corresponding target logical label is obtained from the final logical measurement. For logical observable $k$, define
\begin{equation}
\label{eq:ExperimentalLogicalLabel}
    \lambda_k^{\mathrm{exp}}
    =
    b_k^{\mathrm{log}}
    \oplus
    \bigoplus_{a\in \mathcal{L}_k}m_a ,
\end{equation}
where $\mathcal{L}_k$ is the set of final measurement outcomes whose parity gives the tracked logical observable, and $b_k^{\mathrm{log}}$ is the value expected from the logical state prepared by the noiseless circuit. The shot is counted as a logical failure when
\begin{equation}
    \hat{\boldsymbol{\lambda}}(\mathbf{v}^{\mathrm{exp}})
    \neq
    \boldsymbol{\lambda}^{\mathrm{exp}},
\end{equation}
or equivalently when the corrected final logical outcome differs from the logical state prepared by the noiseless circuit. To clarify the difference between them, $\boldsymbol{\lambda}^{\mathrm{DEM}}$ from Eq.~\eqref{eq:DEMSampling} is a model random variable used in DEM sampling and derivations, while $\boldsymbol{\lambda}^{\mathrm{exp}}$ is the experimentally measured logical label used to benchmark the decoder.

The rate-estimation formulas presented in Appendix~\ref{app:Rates} often depend only on the detector support set $E_e$, not explicitly on the logical-observable support set $L_e$. When the logical-observable support is fixed by the chosen reference DEM (e.g., SI1000, RL, or IBM-like), or is irrelevant to a detector-moment calculation, we write $p_E$ as shorthand for the detector-level probability assigned to support set $E$. If multiple DEM events have the same detector support set but different logical-observable support sets, detector moments alone determine only their detector-level combined probability; in this work the logical-observable support is inherited from the chosen reference DEM.

A DEM event is called \emph{graphlike} if its detector support set has size one or two. If $E_e=\{i,j\}$, the event is represented as an edge between two detector vertices, and its probability is denoted $p_{ij}$ below. If $E_e=\{i\}$, the event is represented as an edge to a boundary, and its probability is denoted $p_{ii}$. For an independent graphlike DEM, MWPM decoders assign the log-likelihood weight
\begin{equation}
\label{eq:w_E}
    w_E=\log\frac{1-p_E}{p_E}
\end{equation}
to each edge and search for a likely set of edges whose boundary matches the observed detector-event pattern. This graphlike setting is the natural input for PyMatching and related sparse-blossom implementations \cite{HiggotACM2022,HiggotQuantum2025}.

A DEM event whose detector support set has size three or more is a hyperedge. Hyperedges can arise from correlated faults, circuit-level faults that propagate to multiple detector changes, leakage, crosstalk, coherent errors, or other effects not well represented by independent graphlike Pauli faults. A graphlike decoder can use such events only after an approximation, such as decomposing a hyperedge into graphlike components. This replacement changes the likelihood model because a single correlated event is represented by several graphlike events. More refined decoders, including correlated matching, belief-propagation methods, search-based decoders, or neural decoders, can use more of the hypergraph or correlation structure directly \cite{BeniArXiv2025,mullerArXiv2025,SeniorArXiv2026}.

The distinction between the DEM support and the DEM event/mechanism probabilities is central to our work. By DEM support we mean the collection of allowed detector-logical signatures $(\mathbf{h}_e,\boldsymbol{\ell}_e)$, or equivalently $(E_e,L_e)$, included as DEM events before their probabilities are specified. This support may be generated from a circuit-level stochastic model, for example using Stim \cite{Gidney2021Stim}. The probabilities $p_e$ may then be obtained from hardware calibration, decoder-prior optimization, or direct estimation from syndrome data. A syndrome-estimated DEM therefore does not claim to reconstruct the microscopic device Hamiltonian or the full quantum channel. Rather, it estimates the effective detector-level stochastic model seen by the decoder under the experimental conditions. This is precisely the information needed to assign decoder weights and evaluate how well different DEMs predict logical performance.

\section{Equations for estimated rates \label{app:Rates}}
For completeness, we provide the analytical expressions used to estimate graphlike edge, boundary-edge, and hyperedge probabilities from detector-event data. Throughout this appendix, angle brackets denote empirical averages over experimental shots when applied to data, and the corresponding expectations over DEM samples when deriving the independent-event formulas. We also use
\begin{equation}
\label{eq:sDefinition}
    s_i=(-1)^{v_i}=1-2v_i,
\end{equation}
where $v_i$ denotes the detector-event bit associated with detector $i$, and
\begin{equation}
\label{eq:mADefinition}
    m_A=\left\langle\prod_{i\in A}s_i\right\rangle
\end{equation}
for any nonempty detector set $A$.

The graphlike edge formulas used here were first proven in Ref.~\cite{Spitz2018}. Assuming no hyperedges, let $p_{ij}$ denote the probability assigned to a graphlike DEM event with detector support set $E=\{i,j\}$, and let $p_{ii}$ denote the probability assigned to a boundary event with detector support set $E=\{i\}$. Then the rates of bulk edges are:
\begin{equation} \label{eq:Bulk}
    p_{ij} = \frac{1}{2}-\sqrt{\frac{1}{4}-\frac{\langle v_iv_j\rangle-\langle v_i\rangle\langle v_j\rangle }{1-2(\langle v_i\rangle+\langle v_j\rangle)+4\langle v_iv_j\rangle}}.
\end{equation}
The error rates of boundary edges are similarly given by:
\begin{equation} \label{eq:Boundary}
    p_{ii}=\frac{1}{2} + \frac{\langle v_i\rangle -1/2}{\prod_{j\neq i}(1-2p_{ij})},
\end{equation}
where $\langle v_i\rangle$ is the empirical firing probability of detector $i$, and $\langle v_iv_j\rangle$ is the empirical coincidence probability for detectors $i$ and $j$.  
When hyperedges are present, Eqs.~(\ref{eq:Bulk}) and~(\ref{eq:Boundary}) are no longer valid by themselves, because higher-order events contribute to lower-order moments. In Ref.~\cite{BrownPRA2025}, we developed a semi-analytical method for estimating hyperedge error probabilities and then redefining lower-order edge probabilities accordingly. 
    
The following equations give a compact inclusion-exclusion form of the analytical syndrome-correlation inversion underlying the formulas of Ref.~\cite{WallraffPRR2026}; related DEM-estimation approaches are developed in Refs.~\cite{BrownPRA2025,YoungArxiv2025}. 
Let $\mathcal{E}_{\leq 4}$ denote the assumed set of detector support sets of size at most four, and let $E\in\mathcal{E}_{\leq 4}$ be one such support set. We define 
\begin{equation}
    q_E=1-2p_E ,
\end{equation}
where $p_E$ denotes the detector-level probability assigned to a DEM event with detector support set $E$, with the logical-observable support inherited from the chosen reference DEM when needed. 
Under the independent-event DEM model, restricted to the support set $\mathcal{E}_{\leq 4}$, the detector moments obey
\begin{equation}
    m_A=\prod_{F\in\mathcal{E}_{\leq 4}}q_F^{|A\cap F|\bmod 2}.
\end{equation}
For a support $E$ with $|E|\leq 4$, define
\begin{equation}
    R_E=\prod_{\emptyset\neq A\subseteq E}m_A^{(-1)^{|A|+1}} .
\end{equation}
Then
\begin{equation}
\label{eq:Hyperedges}
    q_E=
    R_E^{1/2^{|E|-1}}
    \prod_{\substack{F\in\mathcal{E}_{\leq 4}\\ E\subsetneq F}}
    q_F^{-1} .
\end{equation}
For example, for a fourth-order support $E=\{i,j,k,l\}$ with no strict supersets in $\mathcal{E}_{\leq 4}$, Eq.~\eqref{eq:Hyperedges} gives
\begin{equation}
\begin{aligned}
q_{ijkl}
&=
\left(
\frac{
m_i m_j m_k m_l
m_{ijk}m_{ijl}m_{ikl}m_{jkl}
}{
m_{ij}m_{ik}m_{il}m_{jk}m_{jl}m_{kl}m_{ijkl}
}
\right)^{1/8}\\
p_{ijkl}&=\frac{1-q_{ijkl}}{2}.
\end{aligned}
\end{equation}
Similarly, for a third-order support $E=\{i,j,k\}$,
\begin{equation}
\begin{aligned}
q_{ijk}
&=
\left(
\frac{m_i m_j m_k m_{ijk}}
{m_{ij}m_{ik}m_{jk}}
\right)^{1/4}
\prod_{\substack{F\in\mathcal{E}_{\leq 4}\\ \{i,j,k\}\subsetneq F}}
q_F^{-1}\\
p_{ijk}&=\frac{1-q_{ijk}}{2}.
\end{aligned}
\end{equation}
For a graphlike support $E=\{i,j\}$,
\begin{equation}
q_{ij}
=
\left(
\frac{m_i m_j}{m_{ij}}
\right)^{1/2}
\prod_{\substack{F\in\mathcal{E}_{\leq 4}\\ \{i,j\}\subsetneq F}}
q_F^{-1},
\quad
p_{ij}=\frac{1-q_{ij}}{2}.
\end{equation}
Finally, for a boundary support $E=\{i\}$,
\begin{equation}
q_i
=
m_i
\prod_{\substack{F\in\mathcal{E}_{\leq 4}\\ \{i\}\subsetneq F}}
q_F^{-1},
\quad
p_{ii}=\frac{1-q_i}{2}.
\end{equation}
The rates in Eq.~(\ref{eq:Hyperedges}) are estimated hierarchically from $|E|=4$ down to $|E|=1$. Thus, fourth-order rates are estimated first, then accounted for when estimating third-order rates, and so on down to graphlike and boundary rates. For $|E|=4$, the product over strict supersets $F$ is empty.

The reason Eq.~\eqref{eq:Hyperedges} works is the following. In the product defining $R_E$, a detector support set $F$ contributes only through its overlap with subsets $A\subseteq E$. The alternating product cancels all contributions except those from detector support sets $F$ that contain $E$. If $F$ contains $E$, its contribution to $R_E$ is $q_F^{2^{|E|-1}}$. Therefore, $R_E^{1/2^{|E|-1}}$ gives the product of $q_F$ over all modeled supports $F$ containing $E$. Dividing by the already-estimated strict supersets $F\supsetneq E$ yields $q_E$, and hence $p_E=(1-q_E)/2$.

To estimate DEMs for $r=1$, where no hyperedges are modeled, we use Eqs.~(\ref{eq:Bulk}) and~(\ref{eq:Boundary}). For $r>1$, where hyperedges are included, we use the hierarchical inversion in Eq.~(\ref{eq:Hyperedges}). 

The above formulas are valid when the moment combinations used to define $R_E$ lead to real values of $q_E$ and physical probabilities $0\leq p_E\leq 1$. In our data, this condition is satisfied over most of the DEM. When a detector fires with probability greater than $0.5$, the corresponding $\langle s_i \rangle$ is negative, and higher-order $s$ correlators can also be negative. To assess whether the sign of a negative correlator is resolved by the data, we bootstrap the detection-event data by resampling the shots $100$ times. This gives a bootstrap mean $\bar{s}$ and standard deviation $\sigma_s$. If $\bar{s}<0$ 
and $|\bar{s}|/\sigma_s<\mathrm{SNR}$, with $\mathrm{SNR}=0.5$, we treat the sign as unresolved and replace the correlator by a positive regularization floor $\sigma_s$ for the purpose of rate estimation. If $\bar{s}<0$ and $|\bar{s}|/\sigma_s\geq \mathrm{SNR}$, we keep the negative value and proceed. If the resulting root argument is negative, or if the final probability is outside the physical interval, we set the corresponding DEM event probability to zero. This procedure also acts as a diagnostic: it identifies parts of the device for which the assumed independent DEM-event model and support are not adequate. As shown in the main text, even this partial DEM information can improve logical performance relative to the baseline DEMs considered here.

Finally, in the results reported here, we average the correlators in Eqs.~(\ref{eq:Bulk}),~(\ref{eq:Boundary}), and~(\ref{eq:Hyperedges}) over shots only, not over time-translated 
detector locations. Thus, the estimated rates are allowed to vary from one syndrome-extraction cycle to the next. A time-averaged DEM could instead be obtained by aggregating detection events associated with time-translated DEM events before estimating the rates. Another possibility is to estimate rates for a restricted set of cycles, translate those rates to later cycles, and use a separately estimated terminal cycle to account for boundary effects.

\begin{figure*}[!htbp]
    \centering
    \includegraphics[scale=0.76]{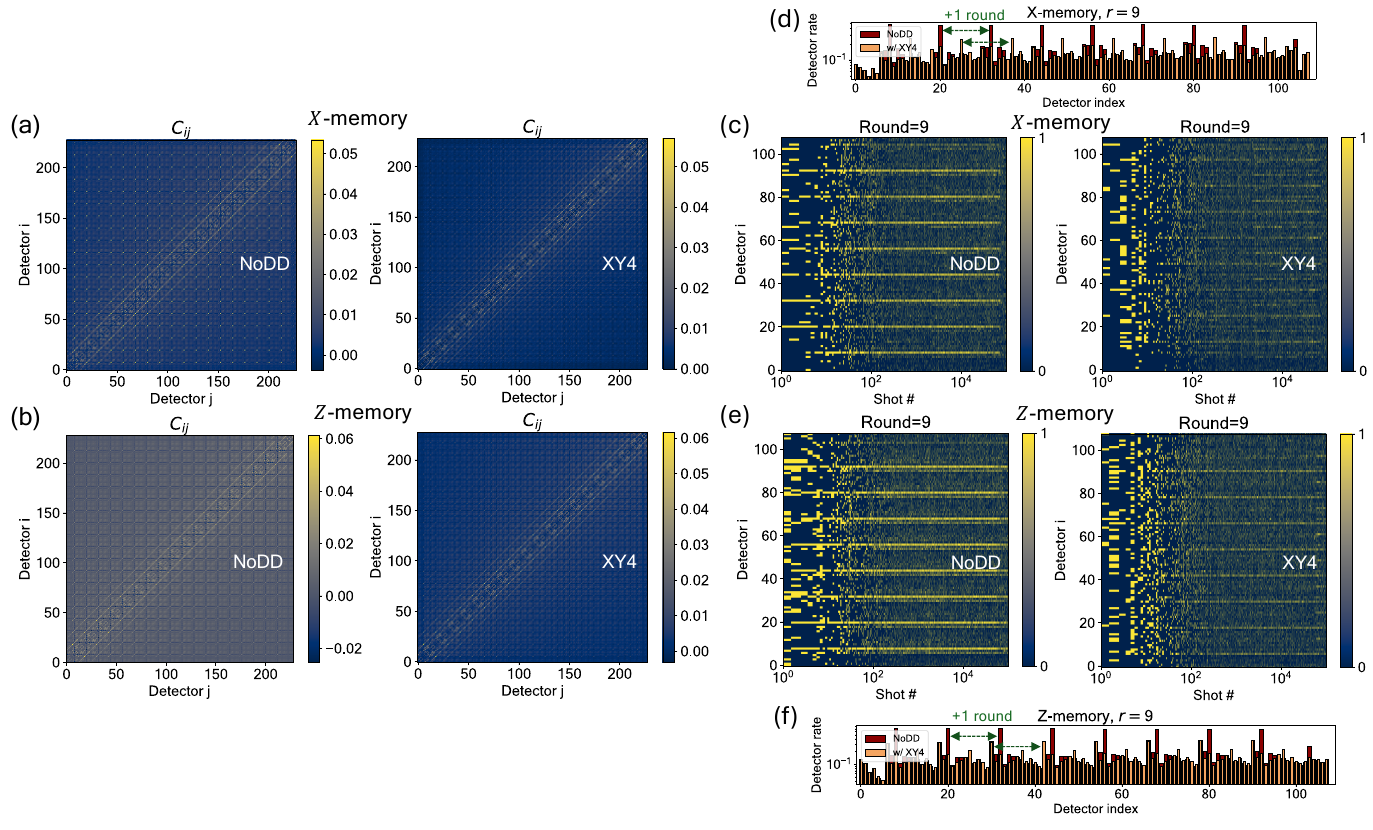}
    \caption{Detector correlations and detector-event structure for \texttt{ibm\_miami}. (a),(b) Detector-pair covariances $C_{ij}=\langle v_iv_j\rangle -\langle v_i\rangle \langle v_j\rangle$ for (a) the $X$- and (b) the $Z$-memory $d=3$ experiments at $r=19$, shown without DD and with XY4. The $Z$-memory without DD exhibits negative covariances, consistent with the negative-correlator issue discussed in the text. The detector index is a flattened space-time index over all $12$ detector locations in all $19$ syndrome-extraction cycles, hence $i\in\{1,\dots,228\}$. 
 (c),(e) Shot-resolved detector-event patterns for the same $X$- and $Z$-memory data for the $r=9$ circuits, using the flattened detector index. (d),(f) Detector firing rates for the corresponding round, comparing NoDD and XY4. 
The green arrows indicate detector locations separated by one syndrome-extraction round.}
    \label{fig:Extra_IBM}
\end{figure*}

\section{IBM\mbox{-}like noise model}\label{app:ibm-like-noise}
To generate the IBM-like reference DEMs for the \texttt{ibm\_miami} experiments, we use a Pauli circuit-level noise model with error rates corresponding to the median qubit performance of the device~\cite{vezvaee2025arxiv}. The physical fault model is:
\begin{itemize}
\item Single-qubit gates: After each single-qubit gate, we apply a single-qubit depolarizing channel with total error probability $p_\mathrm{1q}$.
\item Two-qubit gates: after each two-qubit gate, we apply a two-qubit depolarizing channel with total error probability $p_\mathrm{2q}$.
\item Measurements: we model measurement noise using two independent mechanisms: an $X$ error before measurement and a classical readout flip. Each is applied with probability
\begin{equation}
q_m=1-\sqrt{1-p_m},
\end{equation}
Thus $1-(1-q_m)^2=p_m$ is the probability that at least one of the two measurement-related error mechanisms occurs. [Note that this is not the same as the net probability that the reported measurement bit is flipped, which would be $2q_m(1-q_m)$.]
\item Idle qubits: during an idle interval of duration $t_\mathrm{id}$, we apply a time-dependent biased Pauli channel. To first order in $t_\mathrm{id}$,
\begin{equation}
p_{\mathrm{id},x}=p_{\mathrm{id},y}=\frac{t_\mathrm{id}}{4T_1},\quad p_{\mathrm{id},z}=\frac{t_\mathrm{id}}{2}\left(\frac{1}{T_2}-\frac{1}{2T_1}\right).
\end{equation}
Here $t_\mathrm{id}$ is determined by the compiled circuit schedule, including the median two-qubit and readout durations, $t_\mathrm{2q}$ and $t_m$.
\end{itemize}
This list of circuit-level channels defines a set $\mathcal{F}$ of physical fault locations $f$ in the noisy circuit, with probabilities $r_f$, in the notation of Appendix~\ref{app:DEMIntro}. Each fault $f$ is propagated through the noiseless stabilizer circuit to determine the detector-logical signature $(\mathbf{h}_f,\boldsymbol{\ell}_f)$ that it produces, as defined below Eq.~\eqref{eq:e-map}. The DEM event set $\mathcal{M}$ is then obtained by grouping faults with the same detector-logical signature; the probability of the corresponding DEM event is obtained by the parity rule given in Eq.~\eqref{eq:parity-rule}. The resulting IBM-like DEM supplies both a baseline decoder prior and the reference set of detector-logical signatures used for the syndrome-estimated IBM DEM (denoted $\mathrm{est_{IBM}}$ in the main text). In the latter case, the signatures are kept fixed while the corresponding probabilities are re-estimated from detector moments using the methods of Appendix~\ref{app:Rates}.
The numerical values of $p_\text{1q}$, $p_\text{2q}$, $p_m$, $T_1$, $T_2$, $t_\text{2q}$, and $t_m$ used in the simulations are listed in Table~\ref{tab:ibm_like_parameters}.


\begin{table*}[!htbp]
    \centering
    \renewcommand{\arraystretch}{1.2}
    \begin{tabular}{ccccccc}
        \hline
        $p_{\mathrm{1q}}$ &
        $p_{\mathrm{2q}}$ &
        $p_{\mathrm{m}}$ &
        $T_1\,[\mu\mathrm{s}]$ &
        $T_2\,[\mu\mathrm{s}]$ &
        $t_{\mathrm{2q}}\,[\mathrm{ns}]$ &
        $t_{\mathrm{m}}\,[\mu\mathrm{s}]$ \\
        \hline
        $4.36 \times 10^{-4}$ &
        $1.36 \times 10^{-2}$ &
        $2.2 \times 10^{-2} $ &
        $338.47$ &
        $238.91$ &
        $148$ &
        $2.4$ \\
        \hline
    \end{tabular}
    \caption{Numerical values of the parameters used in the circuit-level model for generating the IBM-like detector error model (DEM).}
    \label{tab:ibm_like_parameters}
\end{table*}

\section{Detector-event diagnostics on \texttt{ibm\_miami} and Google Willow \label{app:ibm-miami_extra}}

Here, we analyze the detector-event data on \texttt{ibm\_miami} in more detail, and compare them with the detector-event data from Google Willow. We begin with \texttt{ibm\_miami}. In Fig.~\ref{fig:Extra_IBM}(a),(b), we show detector-pair covariances $C_{ij}=\langle v_iv_j\rangle - \langle v_i\rangle \langle v_j\rangle$, for both $X$- and $Z$-memory experiments, with and without XY4 DD, at $r=19$ syndrome-extraction cycles. For the $X$-memory without DD, there are several long-range positive covariances, visible as yellow features at detector pairs separated across the grid. These correlations are suppressed under XY4 DD. In the XY4 case, the correlations are more localized in space-time and accumulate mainly near the diagonal region. For the $X$-memory, we also find small negative covariances, which are suppressed and become negligible  when DD is used. For the $Z$-memory without DD, there are more pronounced regions of negative covariance, which are also suppressed under XY4 DD.

\begin{figure*}[!htbp]
    \centering
    \includegraphics[scale=0.776]{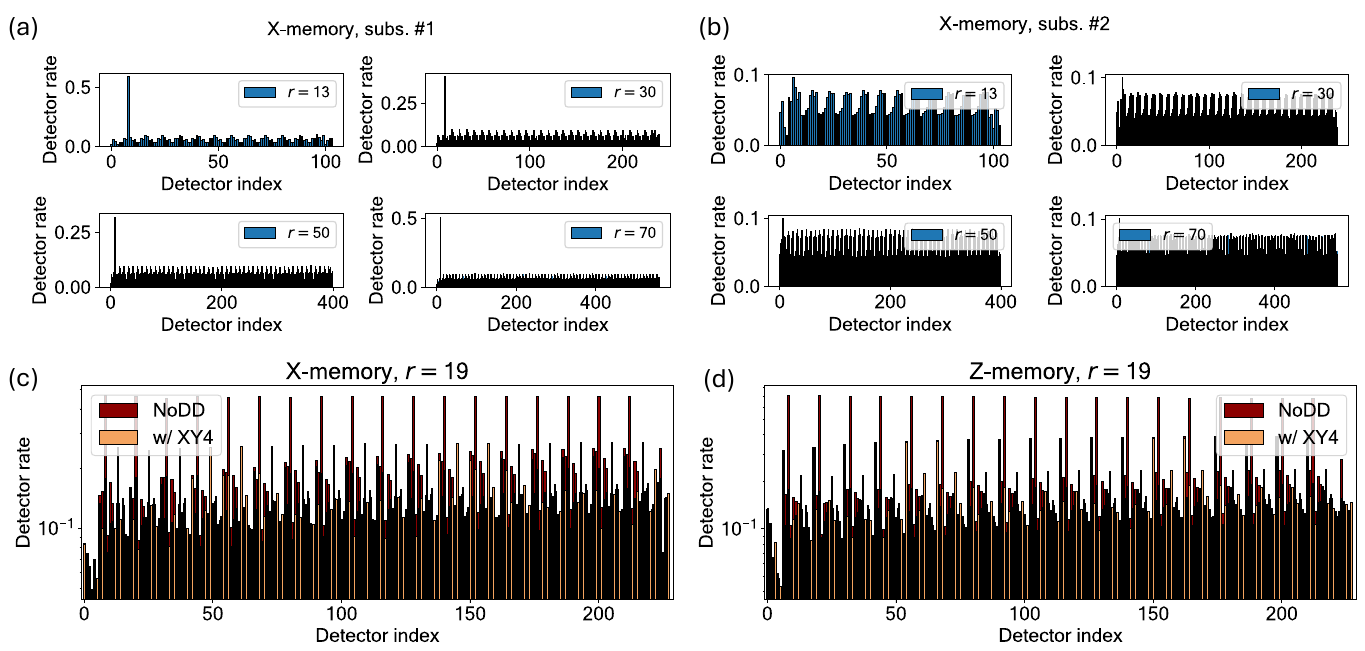}
    \caption{(a) Detector rates over time, for $d=3$ $X$-memory on Google Willow, using subsystem $\#1$. Different panels correspond to $r=13$, $r=30$, $r=50$ and $r=70$ syndrome extraction cycles. (b) Same as in (a), but for subsystem $\#2$. (c) Detector rates over time for $d=3$ $X$-memory on \texttt{ibm\_miami}, and $r=19$ syndrome extraction cycles. (d) Same as in (c), but for $Z$-memory. 
    }
    \label{fig:Det_rates}
\end{figure*}

Another feature we observed is that for both $X$- and $Z$-memory experiments, certain detector events repeat periodically from cycle to cycle with rates substantially higher than those of the remaining detectors. This is illustrated in Fig.~\ref{fig:Extra_IBM}(c), where we show detector events by shot and detector index. The detector index in Fig.~\ref{fig:Extra_IBM} is a flattened space-time index over all detector locations in all syndrome-extraction cycles. Since there are $12$ detectors per cycle, the same detector location in successive cycles appears at indices shifted by $12$. Thus, the horizontal features occur at the time-translated detector family $D_{8+12t}$, with $t=0,1,2,\dots$. The corresponding average detector rates are shown in Fig.~\ref{fig:Extra_IBM}(d). When DD is applied, the most pronounced rates are reduced, although new peaks appear at $D_{13+12t}$, which are also visible in Fig.~\ref{fig:Extra_IBM}(c) for the XY4 case. For the $Z$-memory shown in Fig.~\ref{fig:Extra_IBM}(e), without DD there are two periodic peak families, $D_{8+12t}$ and $D_{6+12t}$. After XY4 DD is applied, the $D_{8+12t}$ family is reduced substantially, while the $D_{6+12t}$ family  remains nearly unchanged. The DD sensitivity of the first family is consistent with a coherent idle-period contribution, such as dephasing or residual ZZ crosstalk between neighboring qubits. 

Such persistent high-rate detection events were observed only in a few cases on the Google data:

\begin{itemize}
    \item A single detector spike on subsystem $\#1$ of the $d=3$ $X$-memory,
    \item A single detector spike on subsystem $\#3$ of the $d=5$ $X$-memory, 
    \item A single detector spike on subsystem $\#3$ of the $d=5$ $Z$-memory,
    \item A single detector spike for the $d=7$ $Z$-memory. 
\end{itemize}
An example of these detector spikes for subsystem $\#1$ is shown in Fig.~\ref{fig:Det_rates}(a). The spike is unlikely to be caused by a shot-local transient, such as a cosmic-ray event, because it appears reproducibly across experiments with different numbers of syndrome-extraction cycles rather than at a random time. Unlike the periodic high-rate detector families observed on \texttt{ibm\_miami}, this high-rate detector does not recur in successive cycles, consistent with the use of reset and leakage-removal mechanisms. For both subsystem $\#1$ and subsystem $\#2$ shown in Fig.~\ref{fig:Det_rates}(a),(b), as well as other subsystems, the detector rates remain approximately constant over time. This behavior is consistent with the leakage-removal techniques used on Google Willow, including data-qubit leakage removal and multi-level reset~\cite{MiaoNatPhys2023}. By contrast, in the \texttt{ibm\_miami} data shown in Figs.~\ref{fig:Det_rates}(c),(d), the detector rates tend to increase slowly with time, apart from the approximately constant periodic peaks identified above. This trend is consistent with leakage accumulation over time, which is plausible because the IBM experiments do not implement leakage removal.

\section{Correlated matching for \texttt{ibm\_miami} \label{ibm_miami_Corr_MWPM}}
\begin{figure}[!t]
    \centering
    \includegraphics[scale=0.8]{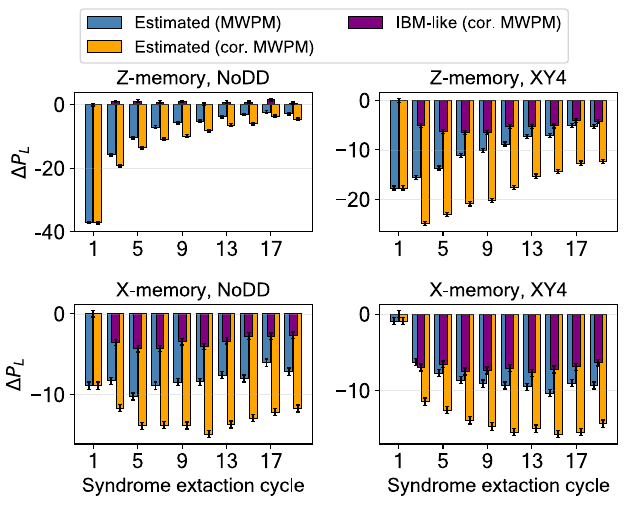}
    \caption{Fractional percent change in LEP relative to the IBM-like DEM decoded with standard MWPM,
$\Delta P_L(A|\mathrm{IBM\mbox{-}like,MWPM})$. The panels show different memory experiments, without DD or with XY4. The blue bars correspond to the estimated DEM decoded with MWPM. The purple bars correspond to the IBM-like DEM decoded with correlated MWPM. The orange bars correspond to the estimated DEM decoded with correlated MWPM.
    }
    \label{fig:corr_Matching_ibm_miami}
\end{figure}

In the main text, we compared the LEPs obtained by decoding with MWPM using either the estimated DEM or the IBM-like DEM. Here we repeat the comparison using correlated MWPM. In Fig.~\ref{fig:corr_Matching_ibm_miami},  all fractional changes are computed relative to the LEP obtained from the IBM-like DEM decoded with standard MWPM, i.e., $\Delta P_L(A|\mathrm{IBM\mbox{-}like,MWPM})$ as defined in Eq.~\eqref{eq:DeltaP_L}.
We evaluate this fractional change for the following cases:
\begin{itemize}
    \item Comparison with estimated DEM decoded with MWPM (blue bars),
    \item Comparison with IBM-like DEM decoded with correlated MWPM (purple bars),
    \item Comparison with estimated DEM decoded with correlated MWPM (orange bars).
\end{itemize}
Negative $\Delta P_L$ indicates improvement relative to the reference value $P_L^{\mathrm{IBM\mbox{-}like, MWPM}}$. The three cases in Fig.~\ref{fig:corr_Matching_ibm_miami} separate the effect of the DEM prior from the effect of the decoder. Comparing the purple and orange bars fixes the decoder to correlated MWPM and changes only the DEM prior: the orange bars, which use the syndrome-estimated DEM, give larger LEP reductions than the purple bars, which use the IBM-like DEM. This shows that the syndrome-estimated DEM provides a more useful detector-level noise model for the decoder. Comparing the blue and purple bars shows that, on the other hand, standard MWPM supplied with the estimated DEM can outperform correlated MWPM supplied with the IBM-like DEM. Thus, improving the detector-level noise model can be at least as important as using a more refined decoder. The best performance is obtained when both improvements are combined, namely when the estimated DEM is decoded with correlated MWPM, given by the orange bars.

For the $Z$-memory without DD, some estimated graphlike rates satisfy $p_E>0.5$, which gives negative log-likelihood matching weights $w_E=\log[(1-p_E)/p_E]$ as defined in Eq.~\eqref{eq:w_E}. Such weights are incompatible with the correlated-MWPM implementation used here. We therefore cap those rates at $0.5$ only for the correlated-MWPM comparison, which sets the corresponding matching weights to zero. This should be interpreted as a decoder-input regularization (not as an estimate of the underlying error mechanism probability). We verified that this regularization does not remove the small LEP improvement obtained for the $Z$-memory without DD.


\section{Hyperedges on Google Willow and \texttt{ibm\_miami}\label{app:Hyperedges}}

\begin{figure*}[!t]
    \centering
    \includegraphics[scale=0.88]{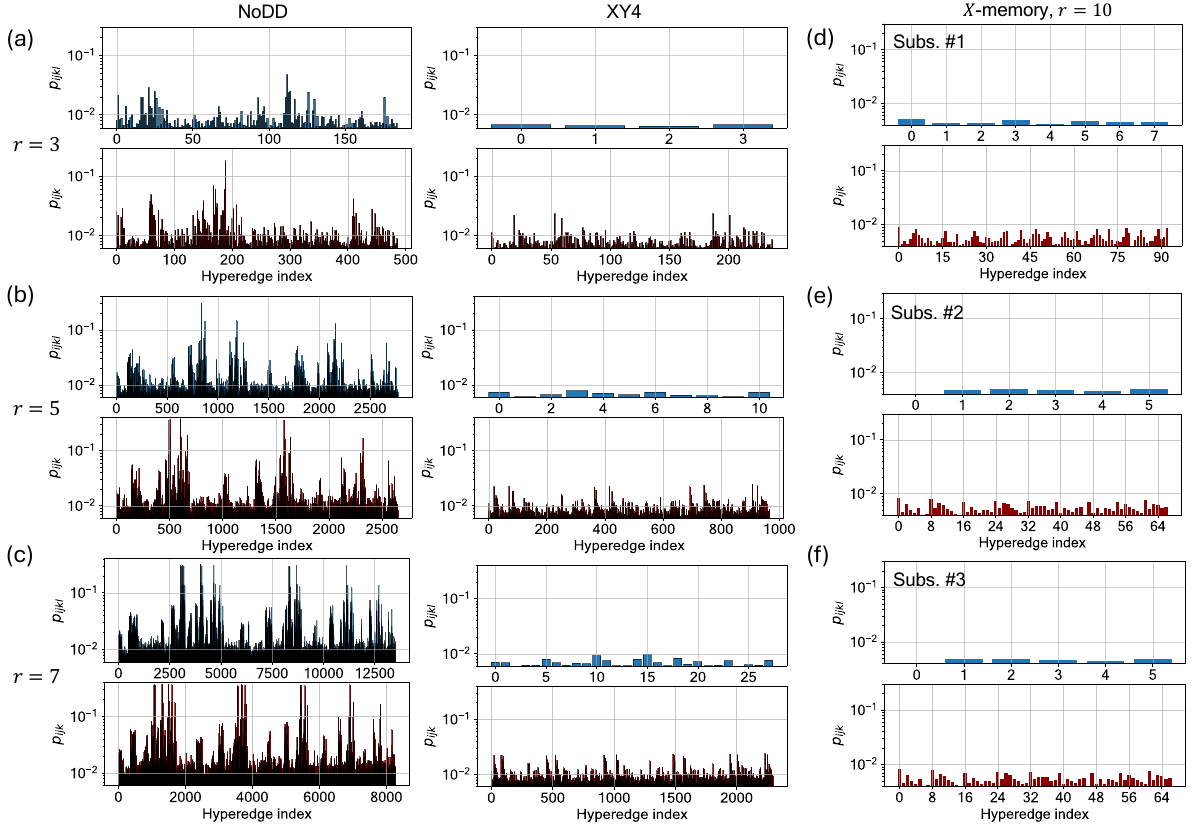}
    \caption{Comparison of effective hyperedge-rate diagnostics between \texttt{ibm\_miami} and Google's Willow chip. (a) Fourth-order diagnostics $p_{ijkl}$ (top) and raw third-order diagnostics $p_{ijk}$ (bottom) for the \texttt{ibm\_miami} $X$-memory experiment at $r=3$, without DD and with XY4. (b) Same as in (a), for $r=5$. (c) Same as in (a), for $r=7$. (d) Fourth-order and raw third-order diagnostics for subsystem $\#1$ of the $d=3$ Willow $X$-memory experiment at $r=10$. (e) Same as in (d), for subsystem $\#2$. (f) Same as in (d), for subsystem $\#3$.
    }
    \label{fig:Hyperedges}
\end{figure*}

As defined in Appendix~\ref{app:DEMIntro}, a hyperedge is a DEM event whose detector support set has size three or more. In this appendix, we use the inversion formulas of Appendix~\ref{app:Rates} as diagnostics for correlated detector structure. Specifically, for candidate detector support sets $E=\{i,j,k\}$ and $E=\{i,j,k,l\}$, we write $p_{ijk}$ and $p_{ijkl}$ as shorthand for the corresponding detector-level quantities $p_E$. These quantities should be interpreted as final DEM probabilities used by the decoder only when the support set is part of the reference DEM support and a corresponding logical-observable support is assigned. Otherwise, they are detector-moment diagnostics. The third-order diagnostics shown here are additionally raw diagnostics: they are not corrected for fourth-order contributions, because we do not assume an independent-event model for arbitrary detector subsets across the full device. 

In Fig.~\ref{fig:Hyperedges}, we evaluate the detector-moment inversion of Appendix~\ref{app:Rates} for candidate detector support sets of sizes four and three in the \texttt{ibm\_miami} and Willow data. Because the IBM experiment uses the unrotated surface code, its DEM contains more detectors than the rotated surface code used for Willow. In Fig.~\ref{fig:Hyperedges}(a), we show the fourth-order and third-order estimates without DD and with XY4. We calculate these quantities across the full detector set, including triplets and quadruplets beyond the detector support sets expected from the stochastic circuit-level noise model generated by Stim. The purpose of this analysis is to diagnose correlated detector structure and possible missing hyperedge signatures. We display only estimates larger than a tolerance, here set to $6\times 10^{-3}$. 

For \texttt{ibm\_miami}, Fig.~\ref{fig:Hyperedges}(a) shows the fourth-order diagnostics and raw third-order diagnostics without DD and with XY4 for $r=3$. Without DD, some raw third-order moment-inversion diagnostics are close to or exceed $0.1$ [see the spikes in Fig.~\ref{fig:Hyperedges}(a) for $p_{ijk}$]. With XY4 DD, the fourth-order diagnostics are suppressed below $10^{-2}$, and the raw third-order diagnostics are suppressed below $2\times 10^{-2}$. As the number of cycles increases to $r=5$ [Fig.~\ref{fig:Hyperedges}(b)] and $r=7$ [Fig.~\ref{fig:Hyperedges}(c)], larger hyperedge diagnostics appear without DD; these are suppressed by XY4 and remain below $2\times 10^{-2}$ in the DD data.

For comparison, we also evaluate the same diagnostics for the $d=3$ Willow $X$-memory rotated surface code at $r=10$ syndrome-extraction cycles. Here we display diagnostic values above $4\times 10^{-3}$. Figure~\ref{fig:Hyperedges}(d) shows the diagnostics for subsystem $\#1$. Compared with \texttt{ibm\_miami}, the Willow estimates are a few times smaller: for example, the raw third-order diagnostics are below $10^{-2}$, compared with values up to about $2\times 10^{-2}$ for \texttt{ibm\_miami} with XY4. We observe a similar behavior in Figs.~\ref{fig:Hyperedges}(e),(f) for subsystems $\#2$ and $\#3$.

\section{Crosstalk test on \texttt{ibm\_miami}}\label{app:Crosstalk}

\begin{figure*}
    \centering
    \includegraphics[scale=0.45]{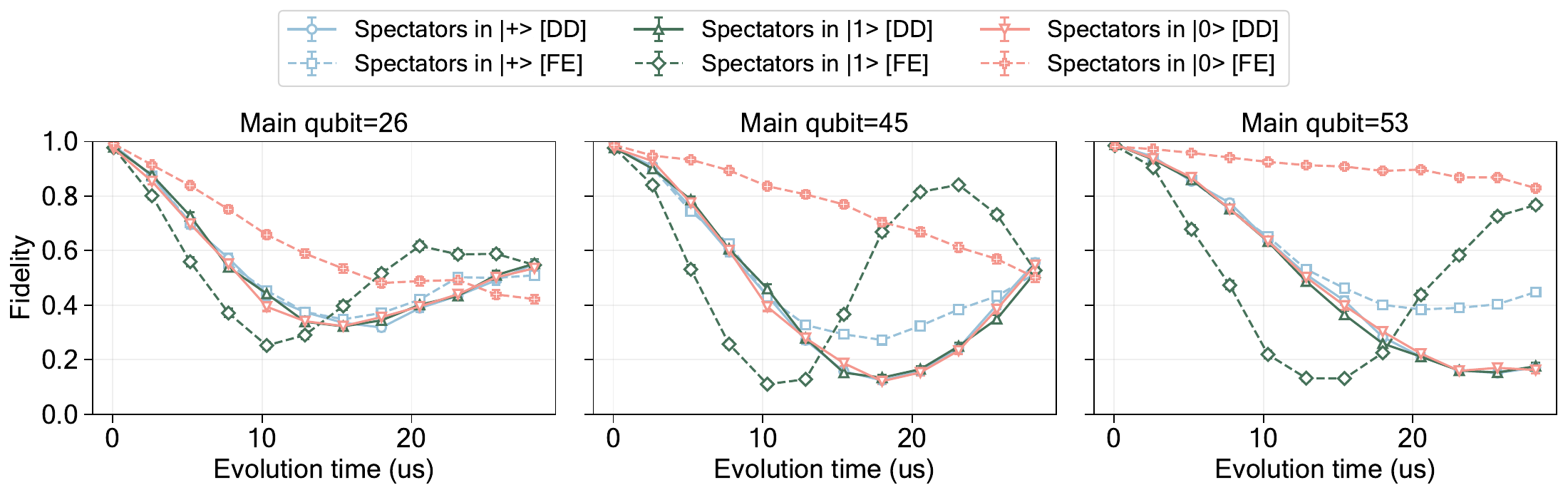}
    \caption{Crosstalk test on \texttt{ibm\_miami}. In each panel, the Main qubit is connected to four adjacent Spectator qubits on the chip: left panel, $26: \{14, 23, 25, 34\}$ \; middle panel, $45: \{35, 44, 46, 55\}$; and right panel, $53: \{43, 52, 54, 63\}$. The dashed lines show the fidelity of the Main qubit as the states of the Spectator qubits are varied. In all panels, changing the Spectator states changes the fidelity of the Main qubit, indicating the presence of crosstalk. Upon applying DD, the crosstalk is suppressed, and the Main-qubit fidelity becomes nearly independent of the Spectator states (solid lines).
    }
    \label{fig:crosstalk}
\end{figure*}

\begin{figure*}
    \centering
    \includegraphics[scale=0.22]{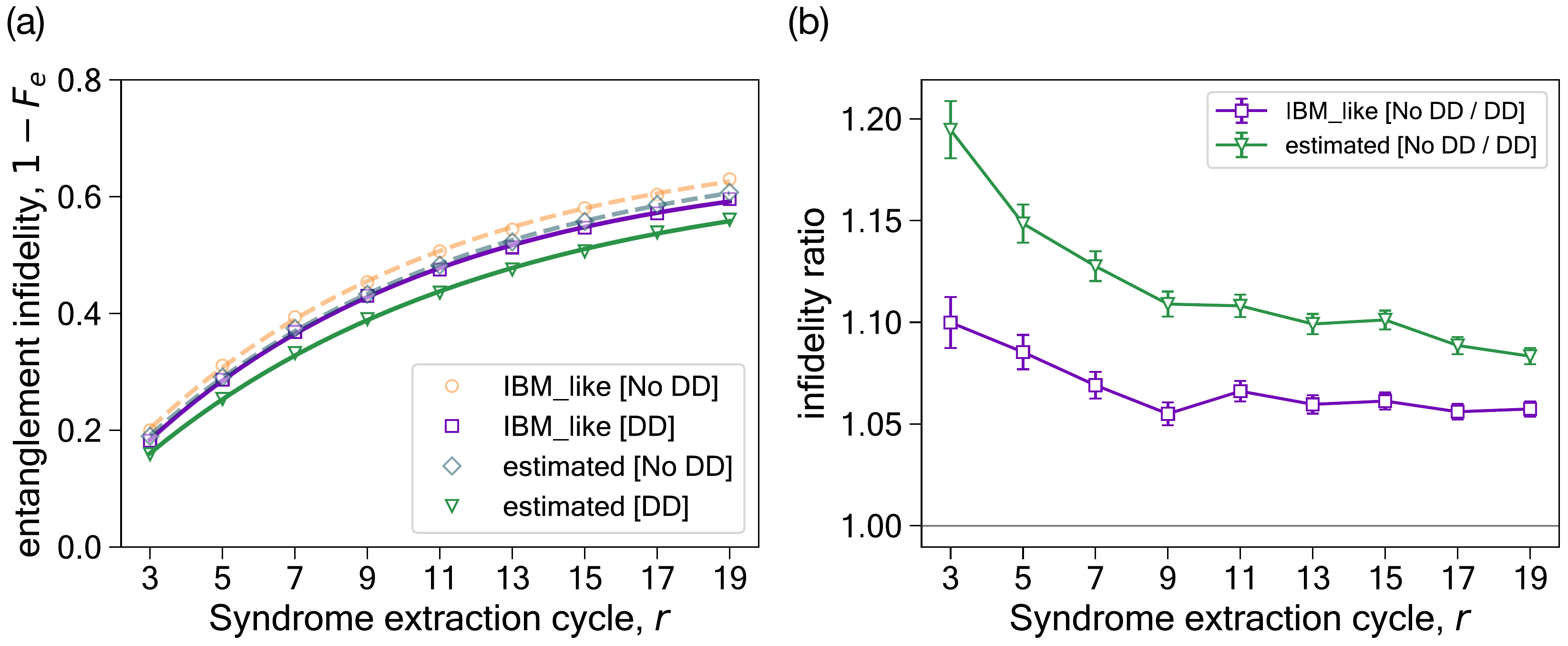}
    \caption{(a) Entanglement infidelity calculated using the IBM-like noise model and the DEM extracted from syndrome data, shown with and without dynamical decoupling. (b) Improvement due to DD, defined as the ratio of the entanglement infidelity without DD to that with DD for a given approach. The syndrome-data-derived DEMs show an additional gain, highlighting the advantage of using DD. 
    }
    \label{fig:ibm-ef}
\end{figure*}

\begin{figure*}
    \centering
    \includegraphics[scale=0.3]{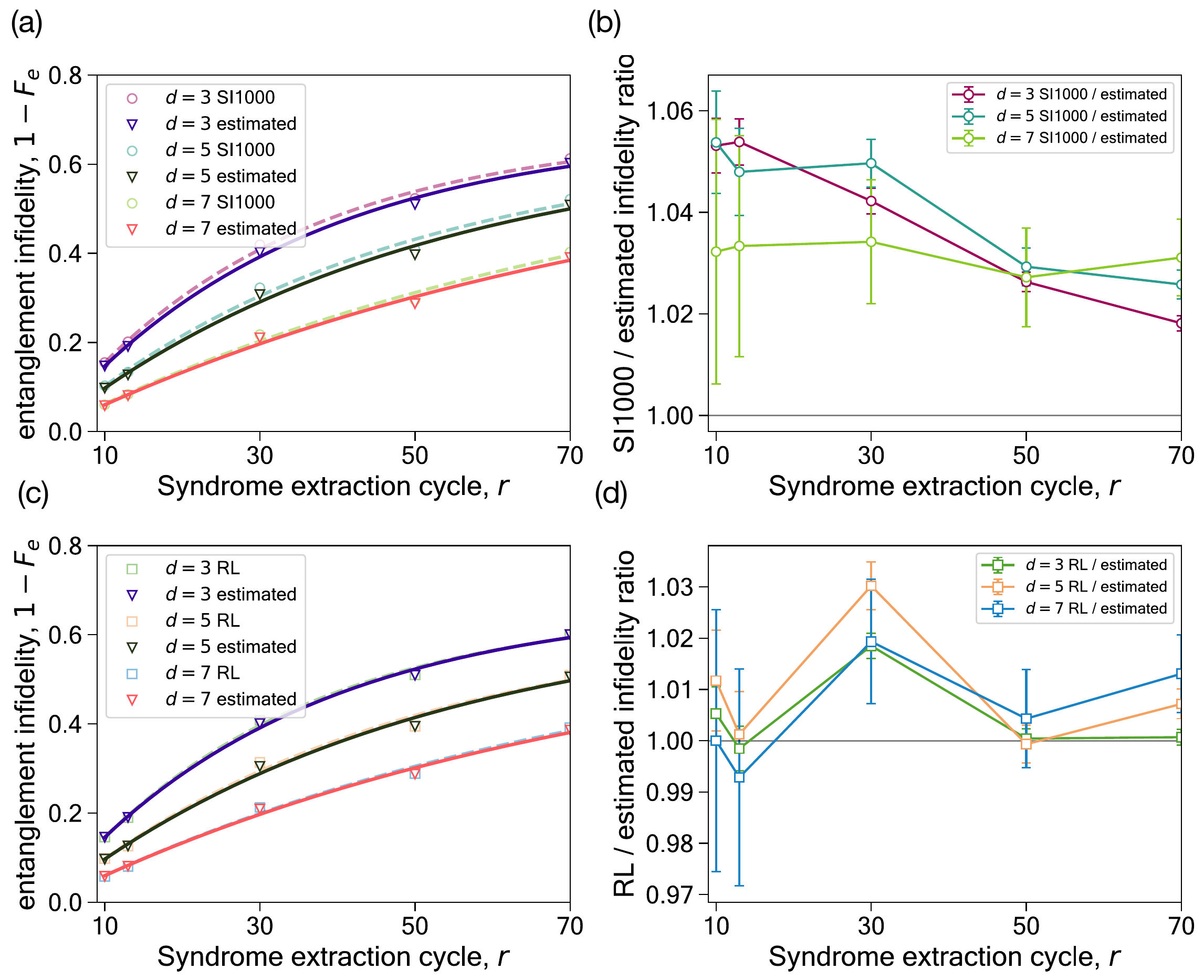}
    \caption{Entanglement infidelity for the Willow code-scaling experiments using the SI1000 noise model (top row) and the RL noise model (bottom row), together with the DEM extracted from syndrome data, shown for code distances $d=3,5,7$. (a) and (c) show the entanglement infidelity for the SI1000 and RL approaches, compared to the DEM extracted from syndrome data for each approach. (b) and (d) show the improvement, defined as the ratio of the entanglement infidelity from the underlying noise model to that from the syndrome-data-derived DEM for a given distance. The syndrome-data-derived DEMs show an additional gain for SI1000, while they are slightly better for RL.
    }
    \label{fig:willow-ef}
\end{figure*}


To assess the presence and severity of crosstalk on \texttt{ibm\_miami}, we perform the same test described in Ref.~\cite{Tripathi2022PRApp}. We prepare a target qubit, labeled the Main qubit, in the $\ket{+}$ state, while preparing  nearby Spectator qubits in one of the states $\ket{0}$, $\ket{1}$, or $\ket{+}$. In the absence of crosstalk, the Main-qubit fidelity should be independent of the Spectator states. As shown in Fig.~\ref{fig:crosstalk}, this is not the case.

Each panel shows a different Main qubit and its neighboring Spectator qubits. The dashed lines show the Main-qubit fidelity over time as the Spectator states are varied. The differences among these fidelities indicate appreciable crosstalk on the device. When staggered DD~\cite{Zeyuan:22,brown2024efficient,evert2024syncopated,Vezvaee2026natcomm} is applied, the crosstalk is suppressed, and the fidelities become nearly independent of the Spectator input states.

\begin{table*}[t]
\centering
\begin{tabular}{lcccc}
\hline
 & RL & estimated RL & SI1000 & estimated SI1000 \\
\hline
$d=3$ & $0.02414 \pm 0.00562$ & $0.02388 \pm 0.00550$ & $0.02552 \pm 0.00641$ & $0.02404 \pm 0.00559$ \\
$d=5$ & $0.01537 \pm 0.00198$ & $0.01510 \pm 0.00187$ & $0.01626 \pm 0.00224$ & $0.01529 \pm 0.00194$ \\
$d=7$ & $0.00901 \pm 0.00067$ & $0.00887 \pm 0.00061$ & $0.00937 \pm 0.00074$ & $0.00895 \pm 0.00065$ \\
\hline
\end{tabular}
\caption{Effective logical error probability per cycle, $\varepsilon$, extracted for the Willow code-scaling experiments by fitting the subset-averaged logical-error-probability curves in the $X$ and $Z$ bases separately, and then averaging the fitted values over the two bases.}
\label{tab:willow_eps}
\end{table*}

\begin{table*}[t]
\centering
\begin{tabular}{lcccc}
\hline
 & RL & estimated RL & SI1000 & estimated SI1000 \\
\hline
$\Lambda_{3/5}$ & $1.570 \pm 0.418$ & $1.582 \pm 0.414$ & $1.569 \pm 0.449$ & $1.572 \pm 0.416$ \\
$\Lambda_{5/7}$ & $1.707 \pm 0.253$ & $1.703 \pm 0.242$ & $1.736 \pm 0.276$ & $1.709 \pm 0.250$ \\
\hline
\end{tabular}
\caption{Suppression factors $\Lambda_{3/5}=\varepsilon_{d=3}/\varepsilon_{d=5}$ and $\Lambda_{5/7}=\varepsilon_{d=5}/\varepsilon_{d=7}$ extracted from the fitted $\varepsilon$ values for the Willow code-scaling experiments.}
\label{tab:willow_lambda_from_eps}
\end{table*}

\section{Entanglement fidelity analysis} \label{app-EF}

Memory experiments consist of preparing the code in a logical state and subjecting it to multiple cycles of stabilizer measurements. In this work, we consider experiments initialized in the logical states $\overline{\ket{+}}$ and $\overline{\ket{0}}$. These experiments directly characterize the code performance only for these two input states. To assess performance for arbitrary logical input states, we use the logical entanglement fidelity (EF) metric developed in Ref.~\cite{vezvaee2025arxiv}.

Since our data include only the $\overline{\ket{+}}$ and $\overline{\ket{0}}$ memory experiments, we use the the SPAM-normalized Pauli lower bound on the EF,
\begin{equation}
F^d_{e}(r)
=
\frac{1}{4}
\left(
1+\frac{\Sigma^d_x(r)}{\overline{\Sigma^d_x}}
\right)
\left(
1+\frac{\Sigma^d_z(r)}{\overline{\Sigma^d_z}}
\right),
\end{equation}
where
\begin{equation}
\Sigma^d_x(r)=1-p^d_{r,+}-p^d_{r,-},
\quad
\Sigma^d_z(r)=1-p^d_{r,0}-p^d_{r,1}.
\end{equation}
Here, $p^d_{r,\alpha}$ denotes the logical error probability after $r$ QEC cycles for code distance $d$ and logical input state $\alpha$. The quantities $\overline{\Sigma^d_{x}}$  and $\overline{\Sigma^d_{z}}$ correspond to first cycle $r=1$ and provide the SPAM renormalization. Because only the $\overline{\ket{+}}$ and $\overline{\ket{0}}$ states are available in our experiments, we take
\begin{equation}
p^d_{r,-}=p^d_{r,+},
\quad
p^d_{r,1}=p^d_{r,0},
\end{equation}
so that the EF lower bound can be computed directly from the measured logical error probabilities. Under the logical-channel assumptions of Ref.~\cite{vezvaee2025arxiv}, and with the two-basis symmetry assumption above for the unavailable conjugate states, this expression gives a SPAM-normalized lower-bound estimate of the entanglement fidelity. Since entanglement fidelity is directly related to the average channel fidelity, it provides a basis-independent figure of merit for the logical memory, rather than a performance estimate tied to a single prepared basis state. This should not be interpreted as a complete model-independent reconstruction of the logical channel, since only the $\overline{\ket{+}}$ and $\overline{\ket{0}}$ memory data are available. Importantly, the estimate depends only on the logical error probabilities output by MWPM decoding with the corresponding DEM, whether from syndrome data or from one of the original approaches.

To provide an additional comparison between DEM extraction from syndrome data and the original methods, we report, for each experimental data set considered here, the resulting EF lower bound computed using both the original DEMs and the syndrome-data-derived DEMs.

Fig.~\ref{fig:ibm-ef} shows the entanglement infidelity, $1-F_\mathrm{e}$, computed from the IBM-like DEM and from the DEM estimated from syndrome data, for experiments performed without DD and with XY4. In both cases, DD reduces the entanglement infidelity across the studied range of syndrome-extraction cycles. Figure~\ref{fig:ibm-ef}(b) quantifies this reduction through the ratio of the entanglement infidelity without DD to that with DD, so that values above unity correspond to improved performance under DD. The syndrome-data-inferred DEM exhibits an additional enhancement, suggesting that it captures DD-induced changes in the effective detector-level noise beyond those predicted by the IBM-like noise model. In Ref.~\cite{vezvaee2025arxiv}, DD performance was optimized using an IBM-like noise model. The results presented here suggest that this optimization could potentially be improved by using a DEM inferred directly from syndrome data, since that DEM more accurately captures the effective detector-level noise seen by the decoder.


Figure~\ref{fig:willow-ef} shows the entanglement infidelity $1-F_\mathrm{e}$ for the Willow code-scaling experiments obtained from the SI1000 and RL noise models and from the corresponding estimates inferred from syndrome data, for code distances $d=3,5,7$. For SI1000, the syndrome-data-derived DEM gives a consistently lower entanglement infidelity over the studied range of syndrome-extraction cycles. For RL, the syndrome-data-derived DEM gives entanglement infidelities that are slightly lower than those obtained from the RL-optimized DEM. These reductions are quantified in Figs.~\ref{fig:willow-ef}(b) and \ref{fig:willow-ef}(d), where we plot the ratio of the entanglement infidelity from the underlying noise model to that from the syndrome-data-derived DEM, so that values above unity indicate an improvement. The effect is present for all three distances in both model classes, but is noticeably stronger for SI1000 than for RL. For RL, the ratios close to unity indicate that the syndrome-data-derived DEM achieves performance comparable to the RL-optimized DEM when the RL-derived detector-logical support and hyperedge decomposition are kept fixed and only the probabilities are replaced by syndrome-estimated values.

Error bars in Figs.~\ref{fig:ibm-ef} and~\ref{fig:willow-ef} are computed by propagating the binomial uncertainties of the logical error probabilities in the $X$ and $Z$ bases, with $\delta p=\sqrt{p(1-p)/N_{\mathrm{shots}}}$, first into $1-F_\mathrm{e}$ and then into the plotted ratios using standard first-order error propagation. The uncertainties of the SPAM-calibration contrasts are not included in this error propagation. For Willow, when plotting subset-averaged quantities, the resulting statistical uncertainties are further divided by $\sqrt{N_{\mathrm{sub}}}$ under the assumption that the subsets are independent. This accounts only for binomial shot noise and does not include possible common-mode device fluctuations or correlations between subsets.

Finally, given that the Willow experiments probe code scaling, it is worthwhile to examine how the syndrome-data-inferred DEMs affect the suppression factor. To do so, we fit the subset-averaged logical error probabilities in the $X$ and $Z$ bases separately to 
\begin{equation}
    p_r=\frac{1}{2}[1-(1-2\varepsilon)^r],
\end{equation}
where $\varepsilon$ is an effective logical error probability per cycle. This fit assumes a time-independent logical failure probability per cycle and no additional offset, and is used here as a phenomenological summary of code scaling. For each code distance and noise description, the logical error data are first averaged over the independent subsets, then the $X$- and $Z$-basis curves are fit independently, and the resulting fitted $\varepsilon$ values are averaged to obtain the reported value. From these fitted $\varepsilon$ values, we define suppression factors $\Lambda_{3/5}=\varepsilon_{d=3}/\varepsilon_{d=5}$ and $\Lambda_{5/7}=\varepsilon_{d=5}/\varepsilon_{d=7}$. The quoted uncertainties on $\varepsilon$ are obtained from the fit covariance for the separate $X$- and $Z$-basis fits and are then propagated through the averaging over bases, while the uncertainties on the suppression factors are obtained by standard propagation of uncertainty for a ratio. The fitted $\varepsilon$ values (Table~\ref{tab:willow_eps}) are consistently lower for the syndrome-data-inferred DEM than for both RL and SI1000 at all distances, indicating improved logical performance per cycle. In contrast, the extracted suppression factors (Table~\ref{tab:willow_lambda_from_eps}) agree within error bars between the baseline and syndrome-data-inferred descriptions. This suggests that syndrome-based estimation improves the overall logical performance without substantially changing the inferred distance scaling;  the suppression factor is therefore less sensitive than the absolute fitted $\varepsilon$ values to this particular decoder-prior improvement.
\clearpage


\bibliography{apssamp}

\end{document}